\documentclass[conference]{IEEEtran}
\usepackage{graphicx}
\usepackage{hyperref}
\usepackage{subfig}
\usepackage{multirow}
\usepackage{amsmath}
\usepackage{lmodern}
\usepackage{booktabs}
\usepackage{url}
\usepackage[table,xcdraw]{xcolor}
\usepackage{algorithm,algorithmic}

\def\BibTeX{{\rm B\kern-.05em{\sc i\kern-.025em b}\kern-.08em
 T\kern-.1667em\lower.7ex\hbox{E}\kern-.125emX}}
\begin{document}
\title{Towards a Supporting Framework for Neuro-Developmental Disorder: Considering Artificial Intelligence, Serious Games and Eye Tracking}
\author{
    \IEEEauthorblockN{Abdul Rehman\textsuperscript{1}, Ilona Heldal\textsuperscript{1}, Diana Stilwell\textsuperscript{2}, and Jerry Chun-Wei Lin\textsuperscript{1}}
\IEEEauthorblockA{
\textsuperscript{1}Department of Computer Science, Electrical Engineering and Mathematical Sciences\\
Western Norway University of Applied Sciences, Bergen, Norway\\
Email: ${\{\text{arj,ilona.heldal,jerry.chun-wei.lin}\}}$@hvl.no  \\ 
\textsuperscript{2}Faculty of Psychology, Alameda da Universidade, 1649-013, Lisbon, Lisbon, Portugal \\
Email: $\text{dianateixeira}$@edu.ulisboa.pt
}
}

\maketitle
\begin{abstract}
This paper focuses on developing a framework for uncovering insights about NDD children's performance (e.g., raw gaze cluster analysis, duration analysis \& area of interest for sustained attention, stimuli expectancy, loss of focus/motivation, inhibitory control) and informing their teachers. The hypothesis behind this work is that self-adaptation of games can contribute to improving students' well-being and performance by suggesting personalized activities (e.g., highlighting stimuli to increase attention or choosing a difficulty level that matches students' abilities). The aim is to examine how AI can be used to help solve this problem. The results would not only contribute to a better understanding of the problems of NDD children and their teachers but also help psychologists to validate the results against their clinical knowledge, improve communication with patients and identify areas for further investigation, e.g., by explaining the decision made and preserving the children's private data in the learning process.
\end{abstract}


\IEEEpeerreviewmaketitle

\section{Introduction}
Neurodevelopmental disorders (NDDs) manifest early in a child's development and affect the growth and function of the nervous system\cite{zavadenko2023neurodevelopmental,gupta2023ai}. These disorders typically affect the structure, function or both of the brain. NDDs are usually diagnosed in childhood and persist in adolescence and adulthood \cite{gogineni2023neurodevelopmental}. Recent technological advances \cite{mumenin2023diagnosis,kollias2022autism} offer promising tools for early detection, support and personalized intervention, validating outcomes, improving clinical assessment, enhancing patient communication, identifying areas for further investigation and creating improvement plans. The combination of AI, serious games and eye-tracking technology can provide a robust framework for understanding and addressing NDDs.

The motivation for this research is rooted in the desire to provide individuals with the best possible support, intervention and care. Using eye-tracking data obtained from computer games, this study aims to propose an approach to support teachers, psychiatrists and other stakeholders by analyzing eye-tracking and serious games-based data to gain insights into student performance. The main objectives of this study are the following.
\begin{itemize}
 \item \textbf{Obj1: }To preprocess data and provide student performance insights (i.e., raw gaze cluster analysis, duration analysis \& area of interest for sustained attention, stimuli expectancy, Loss of focus/motivation, inhibitory control) to inform teachers and to allow games' self-adaptation to improve student well-being and performance by showing personalized actions (i.e., highlighting stimuli to increase student attention). 
 \item \textbf{Obj2: }To predict patterns that are important for teachers and that indicate that the games can improve children's well-being and performance. This should be done by demonstrating personalized actions (i.e., highlighting stimuli to make children more attentive).
 \item \textbf{Obj3: }To provide explainability of the decision made and preserve the private data of the children in the learning progress.
\end{itemize}

\section{Proposed Framework}
\subsection{Previous Work}
Several previous studies have laid the foundation for this research. Costescu et al. \cite{costescu2020development} focused primarily on providing ideas for such a self-adaptive platform. An important idea of such a platform is to determine the focus of attention during children's play. This can be done in the game "Mushroom Hunter" to examine the ability to sustain attention. \cite{bueno2023datasets} focused on the development and application of datasets tailored for AI research in education, especially for children with NDD. Another study \cite{thill2022modelling} explored how robotics and social AI interpret children's behaviour. Another study \cite{daehlen2024towards} explored how serious games combined with eye-tracking technology can provide insights to help teachers better support children with NDD. The study proposed here is based on research on the development of platforms to support NDD children \cite{costescu2020development}, where we focus on three objectives in particular: Processing and analyzing datasets, supporting teachers and psychologists, explainability, and privacy.

\subsection{Framework}
\begin{figure}[!ht]
\centering
\includegraphics[width=0.8\columnwidth]{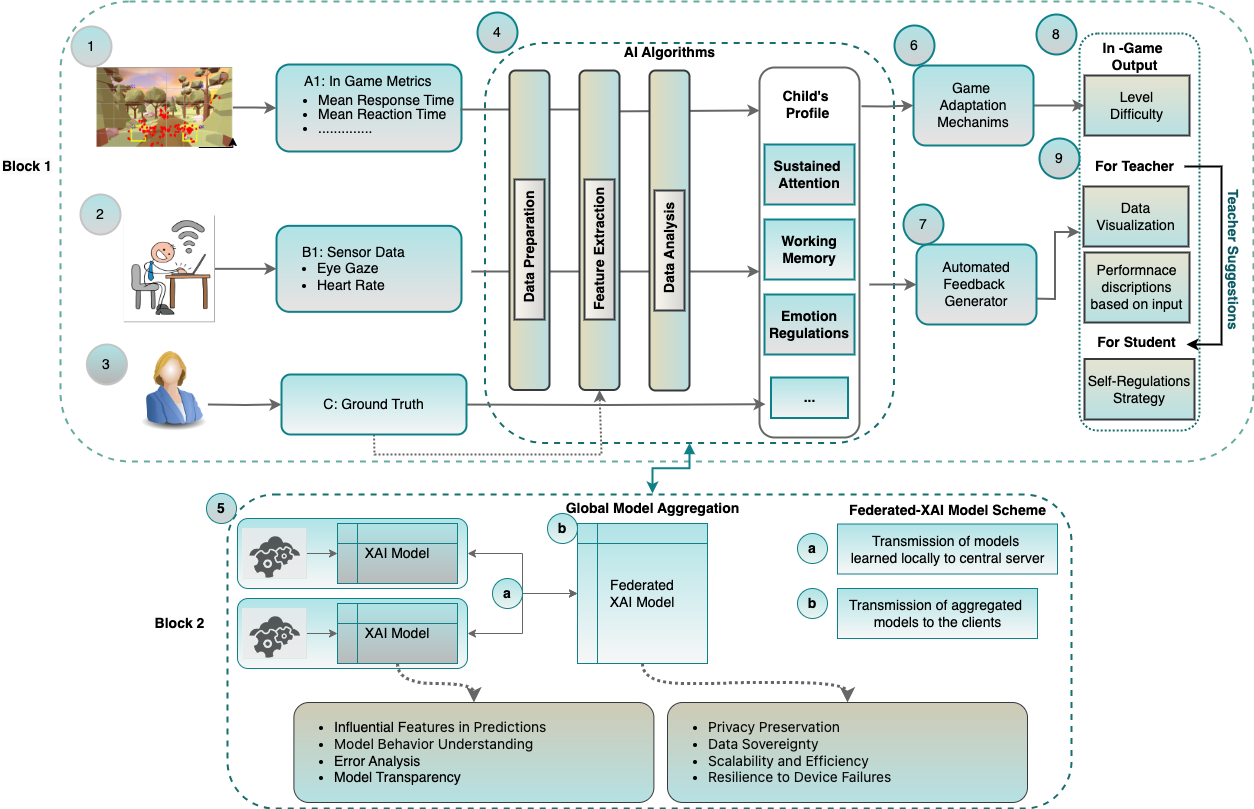}
\caption{Overview of the Proposed Framework}
\label{Overview of the Proposed Framework}
 \end{figure}

Fig. \ref{Overview of the Proposed Framework} shows the flowchart of the proposed framework. The framework consists of 2 main blocks and a total of 9 overall small blocks, where block 1 was taken from \cite{joana}. Currently, this framework is being tested with the game ``Attention'' or ``Mushroom Hunters'', where we are investigating the ability to sustain attention \cite{costescu2023mushroom}.

\begin{itemize}
\item \textbf{Dataset Preparation (Obj1)}:
Currently, the dataset from the game Mushroom Hunter is collected from NDD children using an eye tracker while playing serious games \cite{costescu2023mushroom,daehlen2024towards}. Different use cases will be explored to obtain important information that can help all stakeholders in the project, such as analyzing gaze clusters, duration \& area of interest for sustained attention, stimulus
 expectancy, Loss of focus/motivation, inhibitory control \cite{ali2023towards}.
 \item \textbf{Identification of Changes for Teachers (Obj2):} Assessment and intervention in schools can be personalized to address the needs of the teacher (e.g., critical indicators or mistakes a child makes) and the student (e.g., personalized prompts) to help children in their learning process.
\item \textbf{Explainability and Privacy (Obj3):} A federated explainable AI deep learning model can provide transparent insights into its decision-making process while preserving data privacy across decentralized sources \cite{bucur2024federated}.
 \end{itemize}

\section{Preliminary Findings}	
\begin{figure}[!ht]
\centering
\includegraphics[trim={0 0 0 0 cm},clip,width=\linewidth]{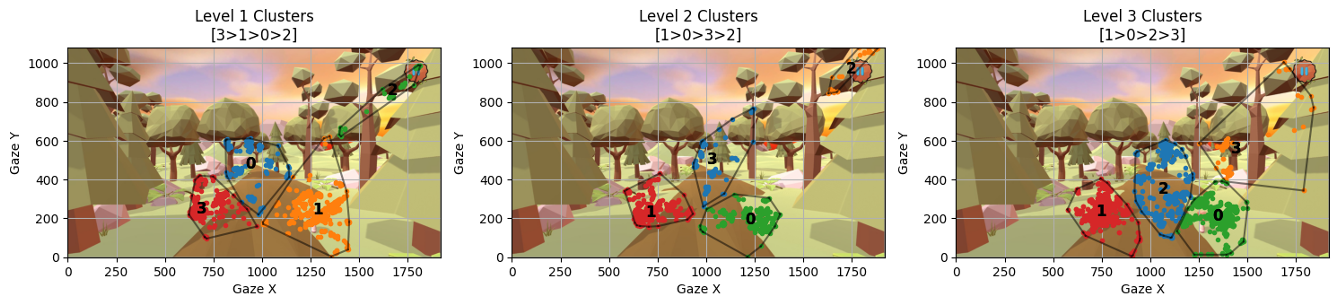}
\caption{Raw Gaze Data Pattern Cluster Analysis.}
\label{rawgaze}
\end{figure}

First, a data analysis is performed, and gaze and fixation patterns are extracted from the raw data. Fig. \ref{rawgaze} shows the four clusters of raw eye gaze data showing the area where the children looked most (i.e. at level 1, the child looked most in cluster 3, then 1, then 0 and then 2) \textbf{(Obj1)}. It should be noted that the child fixated on the left stimuli most of the time, which could indicate limited visual exploration, a preference for specific stimuli, attention deficits and emotional response. This information is provided to teachers and psychologists so that they can support the participant with this problem \textbf{(Obj2)}.

\begin{figure}[!ht]
    \centering
    \subfloat[Duration Analysis\label{duration}]{
        \includegraphics[width=0.46\linewidth]{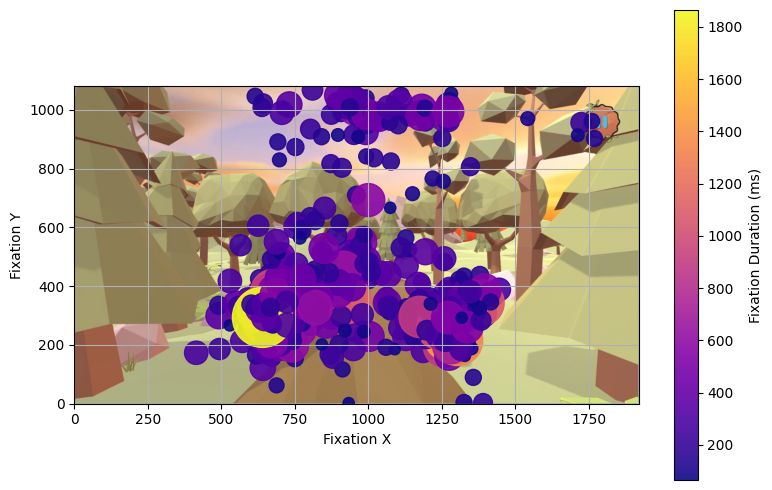}
    }
    \hfill
    \subfloat[Area of Interest Analysis\label{area}]{
        \includegraphics[trim={0 0 0 0.7cm},clip,width=0.45\linewidth]{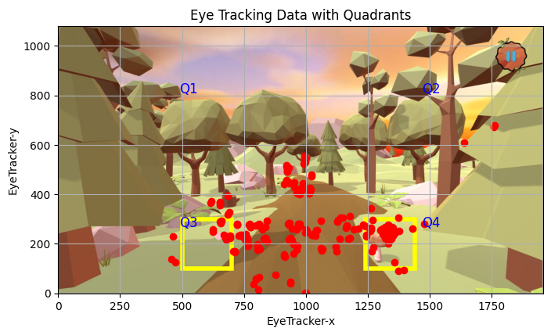}  }
    \caption{Framework Results for Sustained Analysis.}
    \label{fig:attention_analysis}
\end{figure}

Fig. \ref{duration} consists of fixation data in conjunction with the fixation duration. It shows the area on which the child fixated most of the time \textbf{(Obj1)}. It can be seen that the child spent most of the time on the left stimulus. Fig. \ref{area} provides information about whether a child is looking at the area of interest or not \textbf{(Obj1)}. The yellow rectangles are areas of interest. It can be seen that the child fixates most of his time with the right stimuli and has less visual engagement with the other stimuli. This information is made available to teachers and psychologists so that they can support the participant with this problem \textbf{(Obj2)}.

Table \ref{table2} presents the instances when the participant was sustaining attention. The highlighted rows indicate when an object appears, and the participant begins to look at the object in the Q4 quadrant and to the right of the object \textbf{(Obj1)}, which means that the participant was attentive. This information is also provided to teachers and psychologists so that they are aware of this positive behaviour of the participant \textbf{(Obj2)}.

\begin{table}[!ht]
\centering
\caption{Tracking Gaze Data Corresponding to Object Appearance/Disappearance and AoIs.}
\label{table2}
\scalebox{0.75}{
\begin{tabular}{|l|l|l|l|l|l|l|}
\hline
Timestamp & X & Y & Message & Obj-X & Obj-Y & Obj-Z \\ \hline
\rowcolor[HTML]{FFCB2F} 
1.702E+12 & 683 & 319 & Q3-In AoI-Mushroom & 964.4 & 524.4 & -3.5 \\ \hline
1.702E+12 & 842 & 294 & Q3-Not in AoI-Mushroom & & & \\ \hline
1.702E+12 & 1213 & 262 & Q4-AoI-Mushroom & & & \\ \hline
\rowcolor[HTML]{FFCB2F} 
1.702E+12 & 1320 & 224 & Q4-In AoI - No Stimuli & 964.4 & 524.4 & -3.5 \\ \hline
1.702E+12 & 1319 & 225 & Q4-AoI-No Stimuli & & & \\ \hline
\end{tabular}}
\end{table}

\section{Conclusion}
This paper illustrates the main elements of a framework that can be used to identify changes in the performance of NDD children using artificial intelligence. It will support teachers by providing the user's performance assessment and enabling self-adaptation of games to improve students' performance and learning capabilities. It will also explain the decisions made and the privacy of children's data.

\section{Acknowledgment}
The research leading to these results is in the frame of the ``EMPOWER: design and evaluation of technological support tools to empower stakeholders in digital education'' project, which has received funding from the European Union's Horizon Europe programme under grant agreement No 101060918. Views and opinions expressed are, however, those of the authors(s) only and do not necessarily reflect those of the European Union. Neither the European Union nor the granting authority can be held responsible for them.

\bibliographystyle{IEEEtran}
\bibliography{ref}
\end{document}